\documentclass[10pt]{iopart}
\expandafter\let\csname equation*\endcsname\relax
\expandafter\let\csname endequation*\endcsname\relax
\usepackage{amsmath}
\usepackage{amsopn}

\usepackage{booktabs}

\usepackage{verbatim}
\usepackage{hyperref}
\hypersetup{ colorlinks=true, linktoc=all,  linkcolor=blue}
\usepackage{amssymb}

\usepackage{subeqnarray}
\usepackage{caption}
\usepackage{subcaption}
\usepackage{mleftright}
\usepackage{multirow}
\usepackage{lipsum,cite}
\usepackage{amsfonts}
\usepackage{graphicx}
\usepackage{epstopdf}
\usepackage{algorithmic}
\usepackage{algorithm}
\usepackage{soul}
\usepackage[utf8]{inputenc} 
\usepackage[T1]{fontenc}
\usepackage{ulem}
\RequirePackage[mathscr]{eucal}

\def\tablename{Table}

\usepackage{mathrsfs}
\usepackage{mathtools}
\usepackage{cite}
\graphicspath{{figures/}}
\usepackage{url}
\usepackage{tikz}
\usepackage{array}
\usepackage{algorithm,algorithmic}

\newcommand{\TOMOGAN}{TomoGAN}
\newif\iffinal
\iffinal
\newcommand{\zliu}[1]{}
\newcommand{\tekin}[1]{}
\else
\newcommand{\zliu}[1]{{\textcolor{blue}{ Zhengchun: #1 }}}
\newcommand{\tekin}[1]{{\textcolor{red}{ Tekin: #1 }}}
\fi

\def\selin#1{\textcolor{black}{{#1}}}

\newcommand\norm[1]{\left\lVert#1\right\rVert}
\DeclareMathOperator*{\argmax}{arg\,max}
\DeclareMathOperator*{\argmin}{arg\,min}
\DeclareMathOperator{\Rop}{\mathcal{R}}
\DeclareMathOperator{\Fop}{\mathcal{F}}
\DeclareMathOperator{\Hop}{\mathcal{H}}
\DeclareMathOperator{\Gop}{\mathcal{G}}
\DeclareMathOperator{\Kop}{\mathcal{K}}
\DeclareMathOperator{\Nop}{\mathcal{N}}
\DeclareMathOperator{\Qop}{\mathcal{Q}}

\begin{document}
\title[Joint ptycho-tomography with deep generative priors]{Joint ptycho-tomography with deep generative priors}

\author{Selin Aslan$^1$,
        Zhengchun Liu$^2$,
        Viktor Nikitin$^1$,
        Tekin Bicer$^{1,2}$,
        Sven Leyffer$^3$,
        and~Do\u ga~G{\"{u}}rsoy$^{1,4}$}
\address{$^1$ X-ray Science Division, Argonne National Laboratory, 9700 Cass Avenue, Lemont, Illinois 60439, USA (e-mail: selinaslanphd@gmail.com).}
\address{$^2$ Data Science and Learning Division, Argonne National Laboratory, Lemont, IL 60439, USA.}
\address{$^3$ Mathematics and Computer Science Division, Argonne National Laboratory, 9700 Cass Avenue, Lemont, Illinois 60439, USA.}
\address{$^4$ Department of Electrical Engineering and Computer Science, Northwestern University, 2145 Sheridan Road, Evanston, Illinois 60208, USA.}

\begin{abstract}
Joint ptycho-tomography is a powerful computational imaging framework to recover the refractive properties of a 3D object while relaxing the requirements for probe overlap that is common in conventional phase retrieval. We use an augmented Lagrangian scheme for formulating the constrained optimization problem and employ an alternating direction method of multipliers (ADMM) for the joint solution. ADMM allows the problem to be split into smaller and computationally more efficient subproblems: ptychographic phase retrieval, tomographic reconstruction, and regularization of the solution. We extend our ADMM framework with plug-and-play (PnP) denoisers by replacing the regularization subproblem with a general denoising operator based on machine learning. While the PnP framework enables integrating such learned priors as denoising operators, tuning of the denoiser prior remains challenging. To overcome this challenge, we propose a denoiser parameter to control the effect of the denoiser and to accelerate the solution. In our simulations, we demonstrate that our proposed framework with parameter tuning and learned priors generates high-quality reconstructions under limited and noisy measurement data. 
\end{abstract}


\maketitle

\section{Introduction}
\begin{figure}[b]
\centering
\includegraphics[width=.5\textwidth]{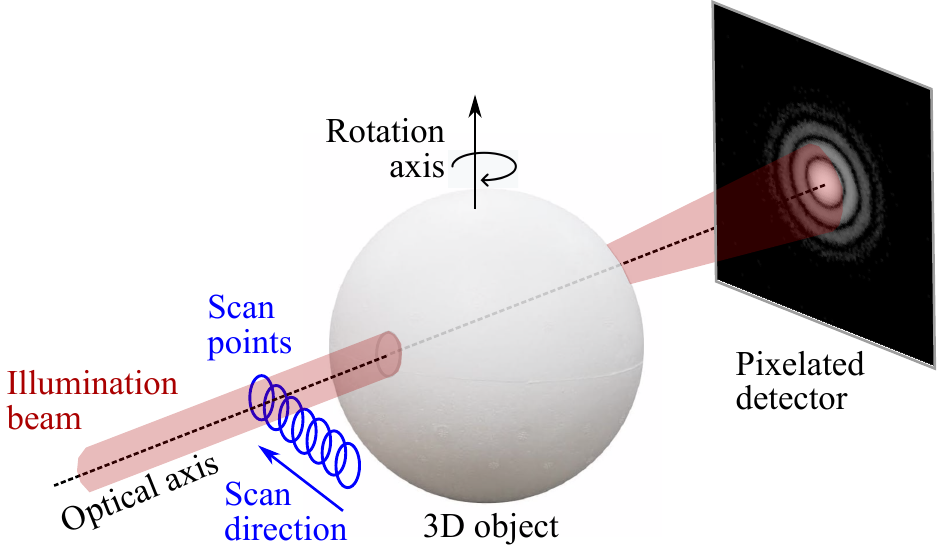}
\caption{Illustration of a ptycho-tomography data acquisition process. A 3D object is scanned with a focused coherent illumination beam while collecting far-field diffraction images with a pixel array detector. This process is repeated for each view of the object around a common rotation axis to collect tomographic data.}
\label{fig:ptycho-tomo}
\end{figure}

Ptychography \cite{Hoppe:69} is a scanning-based coherent diffraction imaging technique that can provide high resolution imaging of thick samples without the need of an optic to form an image. While ptychography can only provide projective imaging of samples in 2D, series of ptychography scans can be acquired in a tomography setting to reconstruct thick volumetric samples in 3D~\cite{Dierolf:10}. In ptycho-tomography, a 3D object is scanned with a small coherent beam to collect a series of diffraction patterns through a pixel array detector located in the far-field; see Fig.~\ref{fig:ptycho-tomo}. The detector records the intensity images of the incident wave on detector plane; therefore, the phase of the wave needs to be recovered through a computational procedure called the phase retrieval. This scanning procedure can be repeated for different view angles of the 3D object around a common rotation axis in order to collect tomographic data and to recover the complex refractive index of the object in 3D. The conventional approach for reconstruction then consists of solving a 2D ptychographic phase retrieval problem independently for each angle, followed by a 3D tomographic reconstruction from the retrieved angular projections of the phase (and amplitude) of the object plane wave. Because phase retrieval algorithms require significant overlap (60$\%$ or more) between neighboring illuminations for a successful recovery, the sequential approach is not optimal and limits scanning large volumes within reasonable data collection times. 

While the sequential approach, that is, first performing phase retrieval for each angle and then tomographically reconstructing the object, is still the method of choice in practice, recent efforts have focused on relaxing or avoiding the illumination overlap requirement. These methods pose the reconstruction problem in a joint fashion. In other words, the phase retrieval problems for each rotation angle are solved simultaneously with the tomographic reconstruction through a joint optimization framework, resulting in a better-posed problem with those extra constraints and allowing for less stringent scanning requirements. Beginning with the first successful demonstration of the joint inversion concept through a numerical simulation \cite{Gursoy:17}, and later on experimentally \cite{kahnt2019coupled}, more efforts have focused on further relaxing these overlap constraints and finding new sparse scanning schemes for high-speed or dose-efficient implementations. Different optimizers such as the Levenberg-Marquardt algorithm \cite{Ramos_19} and Adam algorithm \cite{Du_19} have been used for successfully solving the joint optimization problem. In parallel, an extensible and a generic distributed optimization framework has been proposed in \cite{Aslan:19} as a solution when additional experimental uncertainties due to noise, motion blur, or other types of model mismatches need to be corrected. The framework is based on the ADMM \cite{Boyd:11} and allows splitting the problem into smaller parts where each subproblem can be solved with an independent optimizer. With this modular structure, the whole reconstruction procedure can be expanded by adding new subproblems that often emerge in practical experimental settings. Also, because ADMM sub-problems can be solved independently, we can effectively map those sub-problems onto available computing resources. 

Choosing an appropriate prior for the model is a major challenge for many imaging applications. To tackle this challenge, several regularization methods have been introduced. While some methods define priors explicitly in a regularized optimization framework such as total variation (TV)~\cite{Rudin:92}, Tikhonov regularization~\cite{Tikhonov:77}, and other types of sparsity-based regularization methods~\cite{Tropp:10}; others do not have explicit formulation as an optimization problem, such as BM3D~\cite{Dabov:07} and WNNM~\cite{Gu:14}. We also studied incorporating TV as part of our joint optimization scheme to regularize the solution when data points are significantly reduced or when data is heavily corrupted with noise \cite{Nikitin:19b}. Also recently, learning-based denoisers have been popularized because of their success in improving the quality of low-dose images~\cite{fbpconv}. Unlike physics-based optimization methods, learned priors are based on training a mapping between noisy images and a desirable image, and they are often applied after the reconstruction step is completed \cite{Liu:20} or, in some cases, before the reconstruction in order to improve the raw data \cite{yang2018low}. Furthermore, with the aid of special hardware, the reconstruction times can be improved significantly. One challenge in incorporating learned priors into the ADMM framework is that because the corresponding regularized optimization problem is not explicitly defined, a formal optimization strategy is not applicable. To overcome this challenge, Venkatakrishnan et al.~\cite{Venkatakrishnan:13} proposed the PNP framework, which enables integrating implicit priors for denoising, to enable use of iterative optimization methods. Although the PnP framework was originally proposed ad hoc, it has been popularized quickly in various inverse problems because of its performance \cite{Sreehari:16, Rond:16, Kamilov:17, Buzzard:18, Ye:18, He:19, Sun:19, Wei:20}. This success has also led to related studies; for example, convergence of PnP has also been discussed in studies \cite{Chan:17, Ryu:19, Buzzard:18, Sun:19b}. Another related framework, regularization by denoising (RED), has also been popularized to solve the denoising problem \cite{Romano:17, Metzler:18, Mataev:19}.

\begin{figure}
\centering
\includegraphics[width=0.4\textwidth,]{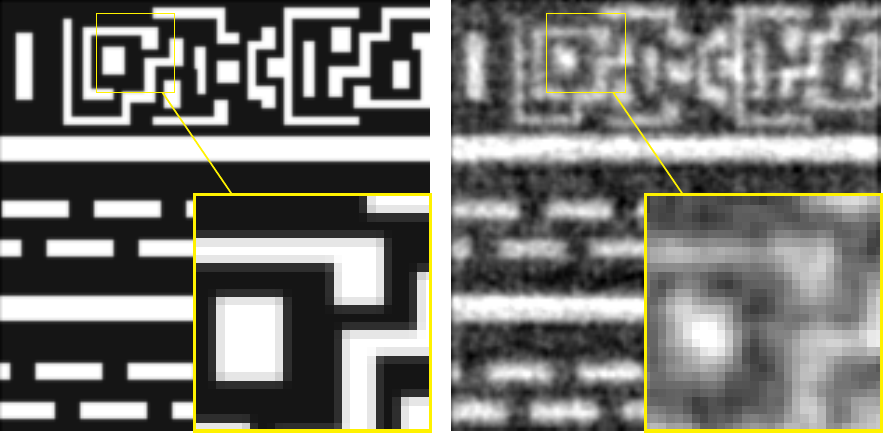}
\caption{Demonstration of the effect of measurement noise on reconstructions: The ground truth (left) and a representative ptycho-tomography reconstruction of a 3D synthetic chip section (right). Because the high-frequency signals in measurement are corrupted more than the low-frequency signals, the resulting effect in the reconstruction is a mix of strong blurring and weak speckle noise.}
\label{fig:Blurr}
\end{figure}

While the PnP framework provides flexible means to incorporate machine-learning-based denoising models into physics-based models, it has been mainly used for additive white Gaussian noise (AWGN) denoising of linear problems.  In ptycho-tomography, or in phase retrieval problems in general, the problem is nonconvex and hard to solve optimally. In addition, the frequency spectrum of reconstruction noise is different from AWGN; see Fig.~\ref{fig:Blurr} for a representative example. This is because the measurements are taken in Fourier space; thus, high-frequency signals dampen quickly, and in turn they are more corrupted than the low-frequency signals because of the Poisson measurement statistics. Therefore, the reconstruction at high measurement noise is blurry, and the state-of-the-art AWGN denoisers are not effective in addressing these types of noise in ptycho-tomography. 

To address the unique challenges of solving ptycho-tomography problem at high noise levels, we propose using pre-trained generative prior models in an ADMM framework. As a generative model, we used conditional coupled generative adversarial network (GAN), however, other types of generative models may also be suitable for the task. We implemented this approach on graphical processing units (GPUs) and validated its effectiveness on realistic data sizes with highly sparse data and noisy measurements. We compare our results with the conventional offline denoising and TV regularization, which are commonly used for denoising in ptycho-tomography applications. Our results show that our optimizations can decrease the total number of required projections (with significantly fewer overlapped regions) by 75\% compared with using adequately sampled data (based on Nyquist) while maintaining good image quality. 

The remainder of this paper is organized as follows. In Section~\ref{sec:backg}, we give an overview of the joint ptycho-tomography problem and its solution using the ADMM method. Section~\ref{sec:prior} describes the challenges in using the original PnP framework and how we tackle the problem. The training, network design, and other important implementation details of the framework are given in Section~\ref{sec:implementation}. In Section~\ref{Sec:NumExp}, we validate our proposed framework for the joint ptycho-tomography problem via simulated experiments. Discussion and conclusions are given in Section~\ref{sec:discussion}.

\section{Background}\label{sec:backg}
In this section, we formulate the ptycho-tomography forward and inverse problems and describe the ADMM scheme for the reconstruction.

\subsection{The forward problem} 
In the ptycho-tomography problem, the model for reconstructing the complex refractive index of a 3D object, $x=\delta +i \beta$, is given by
\begin{equation}\label{Eq:fwd_model}
    \text{Poisson}\{|\Gop\Hop x|^2\} = d.
\end{equation}
Here, we use a Poisson-based measurement model, which accurately captures photon-counting statistics in diffraction data in Fourier space. $\Gop$ is the ptychography operator, $\Hop$ is the tomography operator, $x$ is the unknown object, and $d$ is the measurement data. $\Gop$ is defined as $ \Gop \psi = \Fop \Qop\psi$, where $\psi=\Hop x$ is the object transmission function, $\Fop$ is the discrete Fourier transform operator, and $\Qop$ is the illumination matrix. $\Hop$ is defined as $\Hop x = \exp{({\imath c}\Rop x)}$, where $\imath$ is $\sqrt{-1}$, $c$ is the wavenumber of the illumination beam, and $\Rop$ is the Radon transform \cite{Helgason:99}. 

\subsection{The inverse problem} 
Let $p(x|d)$ be the posterior conditional probability of having an object $x$ with given measurements $d$. Then using Bayes's rule, the maximum a posteriori probability (MAP) estimate for the solution $x_\text{MAP}$ is defined as follows:
\begin{align}\label{Eq:MAPdef}
    x_\text{MAP}&=  \argmax_x \frac{p(d|x) p(x)}{p(d)} \nonumber \\ 
                &= \argmin_x -\log \{p(d|x)\}- \log \{p(x)\},
\end{align}
where $\log p(d|x)$ is the log-likelihood of the observation and $\log p(x)$ is the prior of $x$, also referred to as the regularization term. The MAP estimate in Eq.~\eqref{Eq:MAPdef} for the ptycho-tomography model in Eq.~\eqref{Eq:fwd_model} is given as
\begin{align}\label{Eq:MAP}
    x_\text{MAP}=\argmin_x &\sum_{j=1}^{n}\left(|\Gop\Hop x|_j^2-2d_j\log |\Gop\Hop x|_j\right) +\varphi \Nop(x),
\end{align}
where $\varphi\Nop(x)$ is the regularization term to stabilize or to constrain the solution. For simplicity of notation, $j$ indexes all measurement varieties, namely, detector pixel, rotation angle, and scan position. Next, we rewrite Eq.~\eqref{Eq:MAP} into a consensus form by introducing auxiliary variables $\psi$ and $\eta$:
\begin{equation}\label{Eq:MAP2}
\begin{aligned}
    & \min_{\psi,\eta, x} \sum_{j=1}^{n}\left(|\Gop\psi|_j^2-2d_j\log |\Gop\psi|_j\right)+\varphi \Nop(\eta),\\
    & \text{subject to} \quad 
\begin{cases}
    &\Hop x = \psi,\\ 
    &  x = \eta. 
\end{cases}
\end{aligned}
\end{equation}
The objective function is a real-valued function of complex variables, and its  augmented Lagrangian is a complex-valued function. We follow \cite{Li:15} and work with the following real-valued augmented Lagrangian:
\begin{align}\label{Eq:alagr}
\mathcal{L}_{\rho, \tau}^{ \lambda, \mu} (\psi, x, \eta) =&  \sum_{j=1}^{n} \left(|\Gop\psi|_j^2-2d_j\log|\Gop\psi|_j\right) +  \varphi \Nop(\eta) \nonumber \\ &+2\text{Re}\{\lambda^H(\Hop x-\psi)\}
     +\rho\norm{\Hop x-\psi}_2^2 \\ &+ 2\text{Re}\{\mu^H(x-\eta)\}+ \tau\norm{x-\eta}_2^2,\nonumber
\end{align}
where $\rho>0$ and $\tau>0$ are penalty parameters, $\lambda$ and $\mu$ represent dual variables, and $^H$ corresponds to the Hermitian conjugate. This augmented Lagrangian enables us to include the linear terms, $2\text{Re}\{\lambda^H(\Hop x-\psi)\}$, $\rho\norm{\Hop x-\psi}_2^2$, and $2\text{Re}\{\mu^H(x-\eta)\}$, $\tau\norm{x-\eta}_2^2$ in the L2-terms.

\subsection{Solution to the inverse problem} 

Minimization of Eq.~\eqref{Eq:alagr} can be achieved by ADMM with iteratively solving the sub-problems followed by dual variable updates:
\begin{align}
    \psi^{k+1} = & \argmin_{\psi} \sum_{j=1}^{n} \left(|\Gop\psi|_j^2-2d_j\log|\Gop\psi|_j\right) + \rho\norm{\Hop x^k-\psi+\lambda^k/\rho}_2^2,\label{Eq:ptycho_min}\\
    x^{k+1} = & \argmin_x \rho\norm{\Hop x - \psi^{k+1}+\lambda^k/\rho}_2^2 +  \tau\norm{x -\eta^k+\mu^{k}/\tau}_2^2,\label{Eq:tomo_min} \\ 
    \eta^{k+1} = & \argmin_\eta \varphi \Nop(\eta) +\tau\norm{x^{k+1}-\eta + \mu^{k}/\tau}_2^2,\label{Eq:regul_min} \\
    \lambda^{k+1} = &\lambda^k + \rho \left(\Hop x^{k+1}-\psi^{k+1}\right), \label{Eq:dualvar1}\\
    \mu^{k+1} = & \mu^k + \tau \left(x^{k+1}-\eta^{k+1}\right). \label{Eq:dualvar2}
\end{align}
Using the ADMM framework, we formulate the joint ptycho-tomography problem in Eq.~\eqref{Eq:MAPdef} in terms of three independently defined subproblems: ptychographic phase retrieval in Eq.~\eqref{Eq:ptycho_min}, tomographic reconstruction in Eq.~\eqref{Eq:tomo_min}, and regularization in Eq.~\eqref{Eq:regul_min}. The dual variable updates promote the satisfaction of the constraints in Eq.~\eqref{Eq:dualvar1} and Eq.~\eqref{Eq:dualvar2}.

\subsection{Solutions of the subproblems}
For the first subproblem, we minimize the following objection function: 

\begin{equation}
    F_P(\psi) = \sum_{j=1}^{n} \left(|\Gop\psi|_j^2-2d_j\log|\Gop\psi|_j\right)+ \rho\norm{\Hop x^k-\psi+\lambda^k/\rho}_2^2.
\end{equation}
The corresponding gradient is 
 \begin{equation}
    \nabla_\psi F_P(\psi) =\Gop^H\left( \Gop\psi-\frac{d}{(\Gop\psi)^*}\right) - \rho (\Hop x^k-\psi+\lambda^k/\rho),
\label{Eq:gradptycho}
\end{equation}
which is computed by using the Wirtinger calculus~\cite{Hunger:07}. Here $^*$ denotes the complex conjugate. For the solution, we use the nonlinear conjugate gradient (CG) method \cite{Nocedal:06}:
\begin{equation}\label{eq:CGptych}
    \psi_{m+1} = \psi_m + \gamma_m \xi_m,
\end{equation}
where $\gamma_m$ is a step length computed via a backtracking line search method and $\xi_m$ is the search direction. The first iteration is the steepest descent direction, $\xi_0 = -\nabla_\psi F_P(\psi_0)$. For other iterations, $\xi_{m+1}$ is computed recursively by using the Dai-Yuan~\cite{DaiYuan:99} formula, which gives the fastest convergence in our simulations:
\begin{equation}\label{Eq:DaiYuan}
    \xi_{m+1} = -\nabla_\psi F_P(\psi_{m+1}) + \frac{\|\nabla_\psi F_P(\psi_{m+1})\|_2^2}{y_m^H \xi_{m}} \xi_m,
\end{equation}
where $y_m = (\nabla_\psi F_P(\psi_{m+1}) - \nabla_\psi F_P(\psi_{m}))$. 

For solving the subproblem with respect to $x$ in Eq.~\eqref{Eq:tomo_min}, we transform the nonlinearity introduced by $\Hop x $ as in \cite{Nikitin:19b} and instead minimize the following objection function: 

\begin{equation}\label{Eq:tomo_minCompact}
F_T(x) = \rho\norm{\Kop\Rop x - \zeta}_2^2+\tau\norm{x^{k+1}-\eta +\mu^{k}/\tau}_2^2,
\end{equation}
where the linear diagonal operator $\Kop$ is defined as
\begin{equation}
  \begin{aligned}
    \Kop \Rop x = \frac{2\pi i }{\nu}(\psi^{k+1}-\lambda^k/\rho) \Rop x,
  \end{aligned}
\end{equation}
and $\zeta$ is given by
\begin{equation}
 \zeta = (\psi^{k+1}-\lambda^k/\rho)\log(\psi^{k+1}-\lambda^k/\rho). \\
\end{equation}
Hence, we replace the objective function in Eq.~\eqref{Eq:tomo_min} with Eq.~\eqref{Eq:tomo_minCompact}. The gradient is given as follows:
\begin{equation}
   \nabla_x F_T(x) = \rho\Rop^T\Kop^H(\Kop\Rop x -\xi) + \tau (x-\eta^k +\mu^{k}/\tau). 
\end{equation}
Similar to the ptychography subproblem, we use the CG method with the Dai-Yuan formula; see Eq.~\eqref{eq:CGptych} and Eq.~\eqref{Eq:DaiYuan}.

While Eqs.~\eqref{Eq:ptycho_min}--\eqref{Eq:tomo_min} can be solved via well-known optimization methods, the solution of Eq.~\eqref{Eq:regul_min} depends on the choice of the image prior. The question of how to choose a prior, $-\log \{p(x)\}=\varphi\Nop(\eta)$ is a challenging topic in image processing. While one can choose an explicit image prior and measure its distance using the TV norm, we turn our attention to learning-based priors because of their effectiveness.

\section{Learned priors for denoising}\label{sec:prior} 
In this section, we discuss the solution of the denoising problem. We first rewrite Eq.~\eqref{Eq:regul_min} for some prior $\Nop(\eta)$ as follows:
\begin{equation}\label{Eq:ImgDenoising}
   \eta^{k+1} = \argmin_\eta \Nop(\eta) +\tau/\varphi\norm{\tilde{x}^{k+1}-x}_2^2,
\end{equation}
where $\tilde{x}^{k+1}=x^{k+1} + \mu^{k}/\tau$ and $x$ correspond to the noisy and noise-free images, respectively. Several state-of-the art denoisers do not have closed-form expressions for the prior, $\Nop(\eta)$. Hence, integrating these denoisers into the joint ptycho-tomography problem is challenging. We use the PnP framework~\cite{Venkatakrishnan:13} to replace Eq.~\eqref{Eq:ImgDenoising} with a general denoising operator as follows:
\begin{equation}\label{Eq:PnPdenoiser}
    \eta^{k+1} = \text{Denoiser} \left(\tilde{x}^{k+1} \right),
\end{equation}
where an explicit definition of the image prior, $\Nop(\eta)$, is not necessarily known. While PnP was originally proposed to remove the AWGN of variance, $\sigma^2 = \tau/{2\varphi}$, the method has been extended to Poisson inverse problems \cite{Rond:16}. In this work, we use a Poisson-based MAP model to accurately capture photon-counting statistics in diffraction data. While we still use the ADMM to solve Eq.~\eqref{Eq:MAP2} and while the first two subproblems corresponding to the ptychographic phase retrieval and tomography are the same, the last sub-problem corresponding to the regularization is replaced with a denoising operator in Eq.\eqref{Eq:PnPdenoiser}. The PnP framework allows us to use  state-of-the-art denoising algorithms, such as BM3D~\cite{Dabov:07}, K-SVD~\cite{Elad:06}, and WNNM~\cite{Gu:14}. Although the PnP framework does not give a clear definition of the objective function because of the implicit regularization parameter, the method has shown empirical success in various image reconstruction problems \cite{Rond:16, Kamilov:17, He:19, Wei:20}. Alternatively, deep-learning-based denoisers have shown great success implementing the PnP framework; see, for example, \cite{Zhang:17, RickChang:17, Meinhardt:17}. In this work, we use our recently developed denoising technique based on GAN, whose  implementation details will be discussed in the following section.

We point out that the regularization parameter, $\varphi$, that tunes the regularization term in Eq.~\eqref{Eq:MAP} is associated with the additive noise in the denoising operator, $\sigma^2 = \tau/{2\varphi}$. In our application, we have observed that replacing the regularizer problem, Eq.~\eqref{Eq:ImgDenoising}, by the denoising operator using Eq.~\eqref{Eq:PnPdenoiser} can lead to divergence of the overall ADMM scheme. In particular, it appears that the denoiser pushes early iterations to nonphysical solutions from which the ADMM cannot recover. This observation motivates the introduction of a denoising parameter,  $\alpha^k \in [0,1]$, that controls the influence of the denoising operator. Moreover, ADMM can reach a modest accuracy even when the individual subproblems do not converge to optimal values~\cite{Boyd:11}, and this fact has been used for accelerating ptycho-tomography reconstruction~\cite{Aslan:19}. When acceleration is used, however, the role of the denoiser for approximate solutions of the subproblems needs to be balanced for stabilizing the solution. To this end, we choose a denoiser parameter that gives weight to the data fidelity term at earlier iterations and gradually increases the weight of the denoiser at later iterations as we get closer to the solution. An alternative approach has also been proposed in \cite{Xu:20} when $\Nop(\eta)$ is a closed, convex, and proper function. In particular, we rewrite Eq.~\eqref{Eq:PnPdenoiser} as
\begin{equation}\label{Eq:PnPtunedDenoiser}
    \eta_{k+1} = \alpha^k  \text{Denoiser}(\tilde{x}^{k+1}) +(1-\alpha^k)\tilde{x}^{k+1},
\end{equation}
which makes $\eta_{k+1}$ a convex combination of the denoised reconstructions and the noisy reconstructions, $\tilde{x}^{k+1}$. The extremes $\alpha^k=0$ and $\alpha^k=1$ corresponds to the maximum likelihood (ML) estimate (i.e., no regularizer) and full denoising (i.e., PnP denoiser), respectively. In our implementations, we heuristically choose $\alpha^k$ to provide fast convergence to good reconstructions. 

One challenge that arises from including Eq.~\eqref{Eq:PnPtunedDenoiser} in the ADMM framework is that it does not directly correspond to an optimization problem (unless the denoiser can be written as a gradient) and therefore cannot directly be included in the augmented Lagrangian in Eq.~\eqref{Eq:alagr}. This make it harder to generalize the traditional augmented Lagrangian or ADMM convergence theory.
\begin{figure*}
\centering
\includegraphics[width=.9\textwidth]{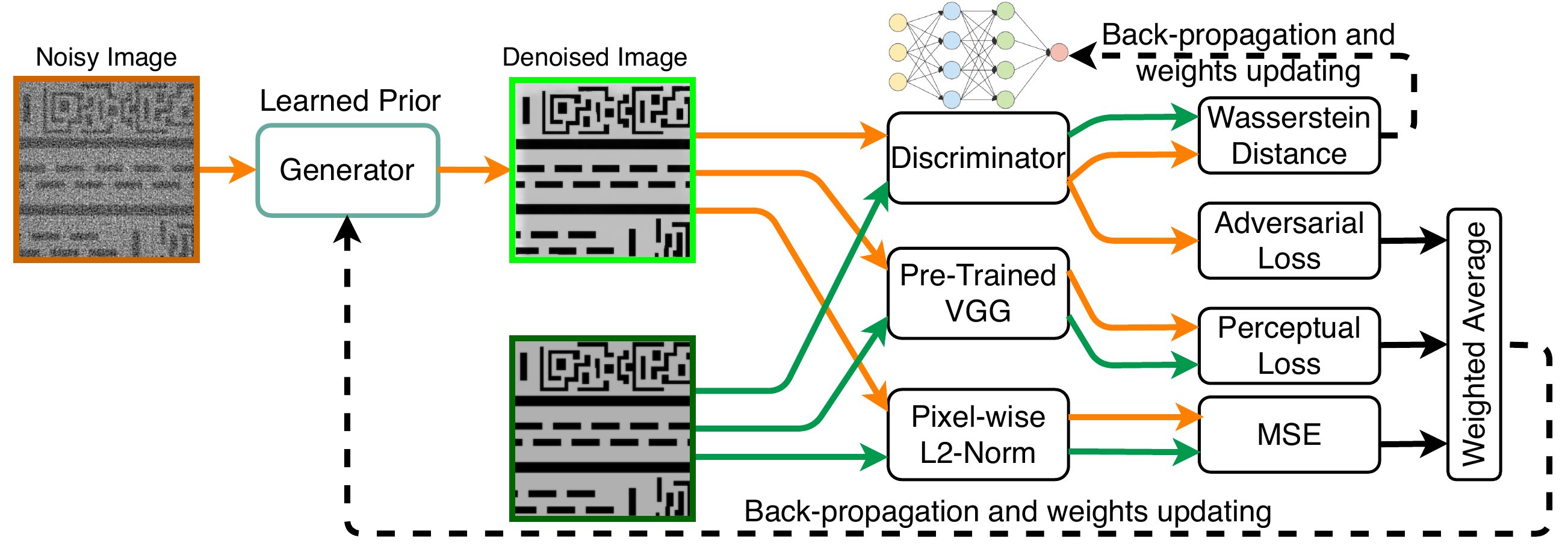}
\caption{Model training pipeline. Once the model is trained, only the generator is used as the learned prior to advance the tomographic reconstructions.}
\label{fig:tomogan}
\end{figure*} 
\section{Implementation}\label{sec:implementation}
In this section, we discuss the implementation aspects of our approach; see Algorithm~\ref{Alg:PnP-ADMM}. For the ptychography and tomography subproblems, we use the same solvers (CG) as in our previous work; see Section 4 in \cite{Nikitin:19b}. Hence, we devote this section to the details of the denoising operator used in Eq.~\eqref{Eq:PnPtunedDenoiser}.

In our simulations, we use our recently developed denoiser, \TOMOGAN{}~\cite{Liu:20}, an image-quality enhancement model based on generative adversarial networks~\cite{2014arXiv1406.2661G}, which was originally developed for low-dose X-ray imaging as the learned prior. Fig. \ref{fig:tomogan} shows the training pipeline of the model where two neural networks (i.e., generator and discriminator) contend with each other during the training until  an equilibrium is reached. 
Specifically, the generative network generates noise-free images from noisy images while the discriminative network evaluates them; thus both networks are trained from the competition. 
The VGG~\cite{vgg} is a neural network model with 19 convolutional neural network (CNN) layers followed by three fully connected layers for image classification. Here, the VGG was pretrained with the ImageNet dataset~\cite{deng2009imagenet}, and we only keep the 19 CNN layers to work as a feature extractor for quantifying the difference between denoised image and true image in VGG's feature space. 
The generator model will work as the learned prior (i.e., for denoising in  Eq.~\eqref{Eq:PnPtunedDenoiser}) once trained by using the pipeline. That is, we can input a noisy image to the generator, and it outputs the corresponding enhanced image. 

The \TOMOGAN{} generator network architecture is a variation of the U-Net architecture proposed for biomedical image segmentation by Shan et al.\cite{8353466}. It comprises a down-sampling network followed by an up-sampling network.
In the down-sampling process, three sets of two convolution kernels (the three boxes) extract feature maps. Then, followed by a pooling layer, the feature map projections are distilled to the most essential elements by using a signal maximizing process.
Ultimately, the feature maps are $1/8$ of the original size.
Successful training should result in the 128 channels in this feature map, retaining important features.
In the up-sampling process, bilinear interpolation is used to expand the feature maps. 
At each layer, high-resolution features from the down-sampling path are concatenated to the up-sampled output from the layer below to form a large number of feature channels.
This structure allows the network to propagate context information to higher-resolution layers, so that the following convolution layer can learn to assemble a more precise output based on this information.
The detailed \TOMOGAN{} generator architecture can be found in ~\cite{Liu:20}.

\begin{figure}
     \begin{subfigure}[b]{0.47\textwidth}
          \includegraphics[width=0.99\textwidth,]{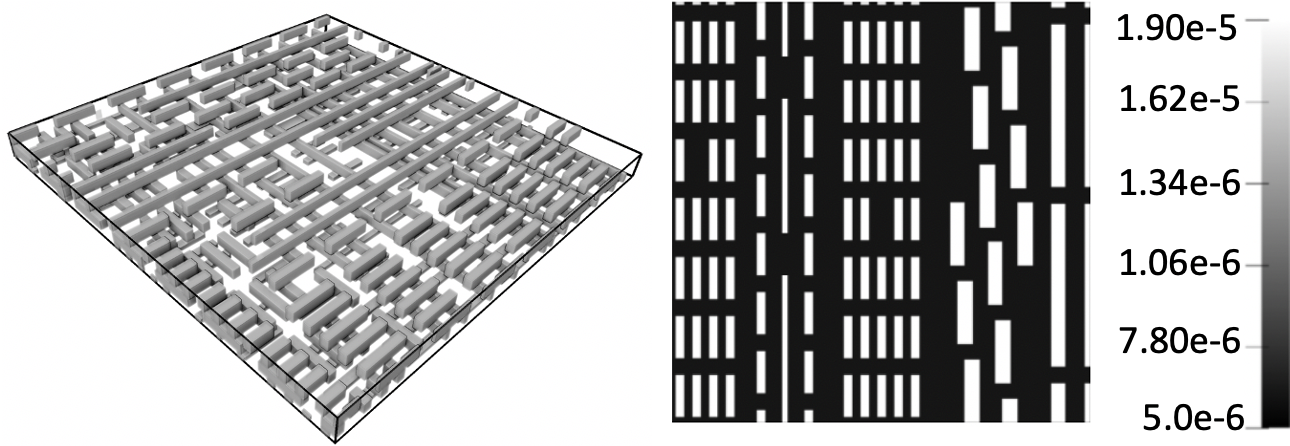}
          \caption{Synthetic chip}
         \label{Fig:3D chip}
     \end{subfigure}
     \begin{subfigure}[b]{0.47\textwidth}
         \includegraphics[width=0.99\textwidth,]{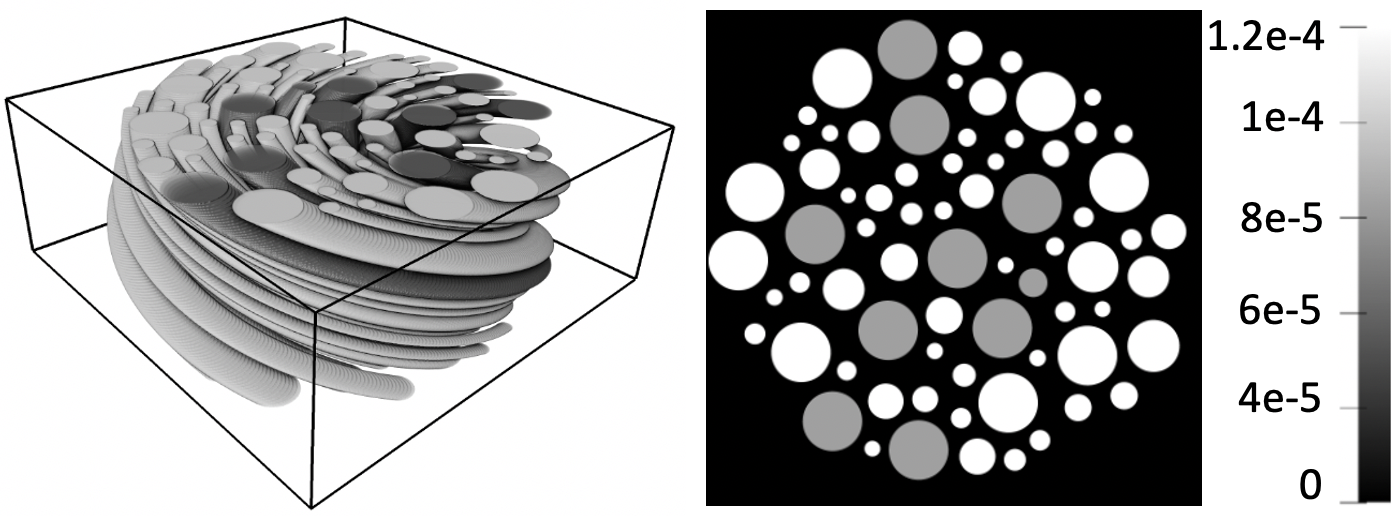}
           \caption{Synthetic phantom}
         \label{Fig:3D phantom}
     \end{subfigure}
     \caption{The 3D objects and corresponding 2D slices. Only the real part of the object ($\delta$) is shown.}
     \label{Fig:3D objects}
\end{figure}

We implemented \TOMOGAN{} with TensorFlow~\cite{abadi2016tensorflow} and used one NVIDIA Tesla V100 GPU card for training \selin{where the total training time is around 6 hours.} The Adam algorithm~\cite{Adam} was used to train both the generator and discriminator, with a batch size of 16 samples. 
In order to train and evaluate the model~(discussed in Section\ref{Sec:NumExp}), we synthesized two different 3D samples as shown in Fig. \ref{Fig:3D objects}.
For each sample, we simulated two different cases with different features~(e.g., removed a few features of the chip, and used different random seeds to construct  circles of the phantom object) for model training and testing separately. As a data augmentation to avoid overfitting, each image of the batch is a patch (of size 128$\times$128) that was randomly cropped from the original 512$\times$512 image~(i.e., slice of the 3D objects). 

\begin{algorithm} 
\caption{Joint ptycho-tomography reconstruction with learned prior}
\begin{algorithmic}
\REQUIRE{Given $0 \leq \alpha^k \leq 1, \rho>0, \tau>0$ and initialize: $\psi^0, \eta^0, x^0, \lambda^0, \mu^0$}
\WHILE{not converged}
\STATE{$\psi^{k+1} \gets \argmin_{\psi} \sum_{j=1}^{n} \left(|\Gop\psi|_j^2-2d_j\log|\Gop\psi|_j\right) + \rho\norm{\Hop x^k-\psi+\lambda^k/\rho}_2^2$}
\STATE{$x^{k+1} \gets \argmin_x \rho\norm{\Hop x-\psi^{k+1}+\lambda^k/\rho}_2^2+ \tau\norm{x-\eta^k+\mu^{k}/\tau}_2^2$}
\STATE{$\tilde{x}^{k+1} \gets x^{k+1} + \mu^{k}/\tau$}
\FOR {$j=1 \cdots M$}
    \STATE{$\eta^{k+1}_j \gets \alpha^k \text{Denoiser}(\tilde{x_j}^{k+1}) +(1-\alpha^k)\tilde{x_j}^{k+1}$}
    \ENDFOR
 \STATE{$\lambda^{k+1} \gets \lambda^k + \rho \left(\Hop x^{k+1}-\psi^{k+1}\right)$}
 \STATE{$\mu^{k+1} \gets \mu^k + \tau \left(x^{k+1}-\eta^{k+1}\right)$}
 \ENDWHILE
\end{algorithmic}
\label{Alg:PnP-ADMM}
\end{algorithm}

\section{Numerical experiments}\label{Sec:NumExp}
In this section, we demonstrate the effectiveness of applying the proposed framework for reconstruction of 3D simulated objects in Fig.~\ref{Fig:3D objects}.
\subsection{Simulation settings}

\begin{figure}
\centering
 \includegraphics[width=0.47\textwidth,]{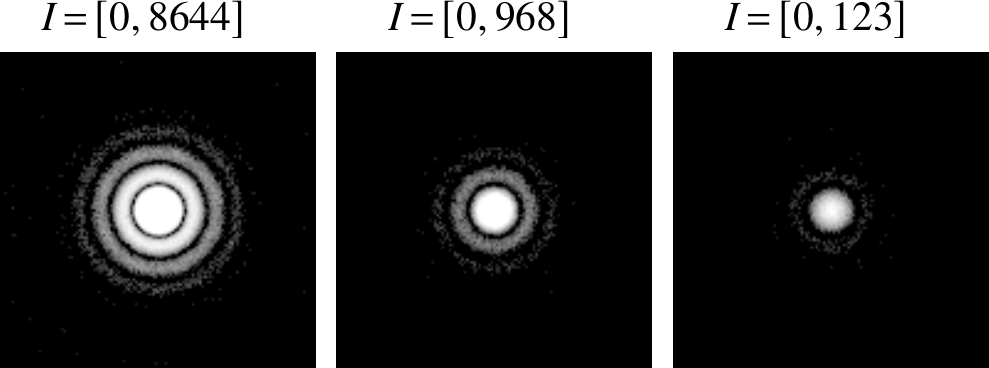}
 \caption{Intensity on the detector for different noise levels. From left to right, the noise level increases.}
 \label{fig:Data}
 \end{figure}

In the first experiment, the object is a simulated chip of size $64\times 512\times 512$ and voxel size 5~nm. The 3D simulated chip and its 2D slice are given in Fig.~\ref{Fig:3D chip}. Our interest is to recover the object that is defined by its complex refractive index, $x=\delta +i \beta$. We use a flat-top Gaussian probe function with probe size $16\times 16$ pixels. The far-field diffraction patterns are recorded by a $128\times 128$ pixelated detector. We use 8.8~keV beam energy to simulate the refractive index values for ptychographic data. We emulate a ptychographic experiment by simulating a 3D chip, where $\delta$ yields the main imaging contrast. We distort the data with Poisson noise. In Fig.~\ref{fig:Data}, we demonstrate the effect of Poisson noise on the measured data for three different detector photon counts in the ranges $I =[0, 8644]$, $I=[0, 968]$, and $I=[0,123]$ on average. As the interval $I$ decreases, the simulations become noisier. 
Initially, the distance between adjacent probe center positions is set to 8 pixels that approximately correspond to $50\%$ overlap. Then, the object is rotated $3N/2$ times at regular intervals from $0$ to $\pi$, satisfying the Nyquist criterion. We refer to this case as \emph{well-sampled data}. Next, we report reconstructions using \emph{under-sampled data} where we further decrease the probe overlap to $25\%$ and rotate the object $3N/8$ times at regular intervals from $0$ to $\pi$. It is essential to point out that solving the ptycho-tomography problem jointly enables relaxing high probe overlap restriction where the conventional methods would fail to reconstruct good quality reconstructions. A more detailed study on conventional methods vs jointly solving the ptycho-tomography problem via ADMM can be found in \cite{Aslan:19}. 

An additional 3D phantom with a shape of $180\times 512\times 512$ and a voxel size 5~nm is generated via XDesign \cite{Ching:17} to demonstrate the effectiveness of the proposed method. The object consists of two different materials: gold (Au), mercury (Hg) with densities of 19.32 g/cm$^3$ and 13.53 g/cm$^3$, respectively. The 3D object and its 2D slice are given in Fig.~\ref{Fig:3D phantom}. While we use 5~keV beam energy to simulate the refractive index values for ptychographic data, the probe and the detector settings are the same as in the first experiment. The detector photon counts are given as  $I=[0, 649]$, and $I=[0,190]$ on average. 

For acceleration of the ADMM, we use 4 inner CG iterations for ptychography and tomography problems, and the ADMM outer iteration limit is set to 250. Early termination is a common practice to accelerate the ADMM solution; see the review in \cite{Boyd:11} and more detailed analysis for our application in~\cite{Aslan:19}. Further accelerations can be possible by varying the penalty parameters $\rho$ and $\tau$ dynamically during the ADMM iterations~\cite[Eq.~(3.13)]{Boyd:11}.
 
\subsection{Simulation results} In this section, we demonstrate the effect of learned priors for the joint ptycho-tomography problem via two 3D simulated objects, see Fig.~\ref{Fig:3D objects}. To quantify image quality degradation, we use the peak signal-to-noise ratio (PSNR). 

 \begin{figure}
     \centering
     \includegraphics[width=0.95\textwidth,]{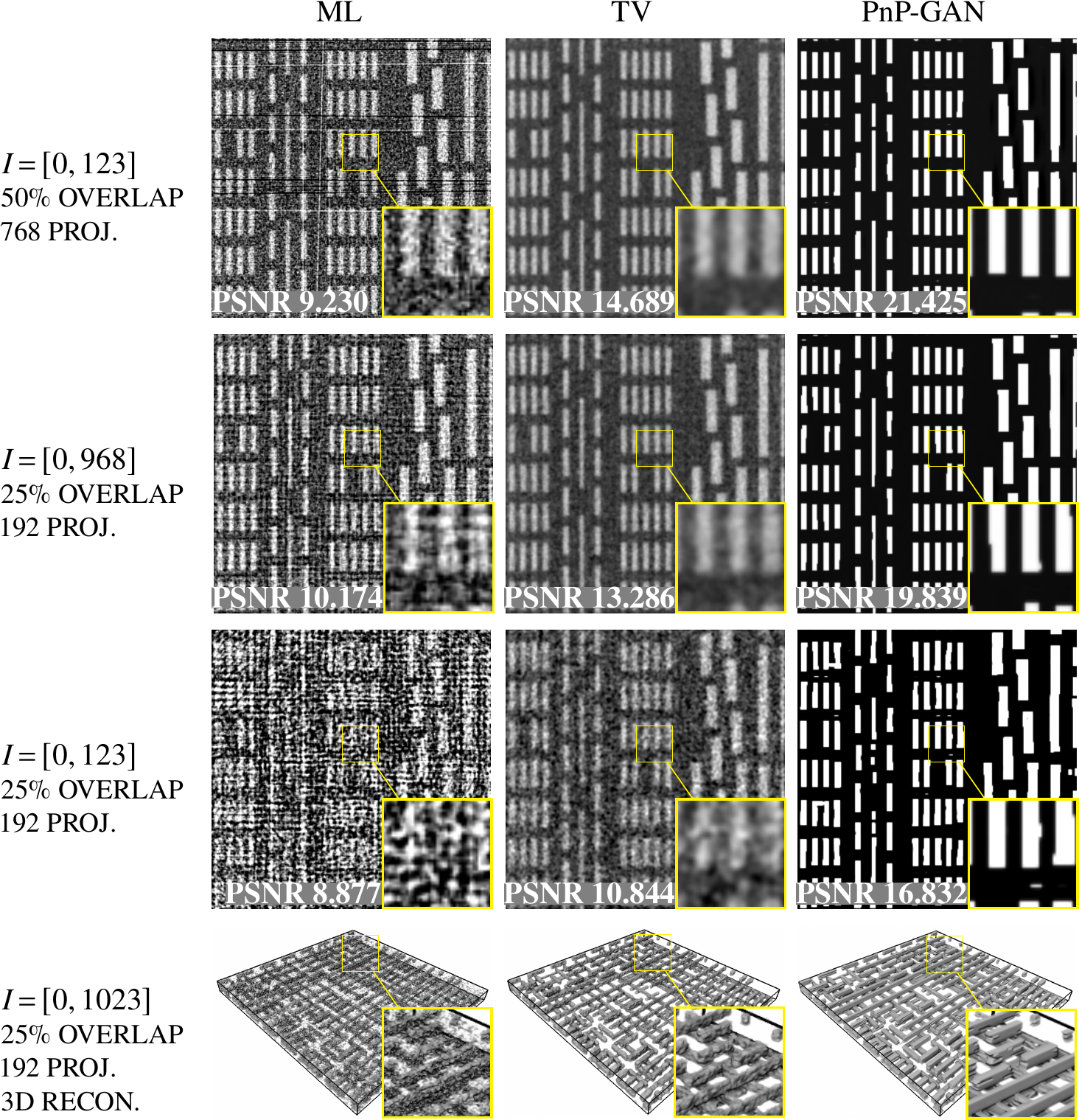}
     \caption{Reconstruction results of a typical slice of the 3D synthetic chip of size $64\times 512\times 512$ at different probe intensity levels. The first row corresponds to reconstructions with $50\%$ probe overlap and 768 projection angles and high noise level. The remaining rows correspond to reconstructions with $25\%$ probe overlap and 192 projection angles and various noise levels. ML: the ADMM outer iteration is 250, and inner CG iterations are 4 for each ptychography and tomography subproblem. TV: the ADMM outer iteration is 250, and inner CG iterations are 4 for each ptychography and tomography subproblem followed by TV subproblem. PnP-GAN: the ADMM outer iteration is 250, and inner CG iterations are set to 4 for each ptychography and tomography subproblem followed by TomoGAN subproblem.}
     \label{fig:Recons}
 \end{figure}
In Fig.~\ref{fig:Recons}, we report reconstruction results for the real part of the object, $\delta$, using three different reconstruction results: (1) the ML estimate (i.e., no regularizer), (2) the MAP estimate with TV prior (i.e. Eq.~\eqref{Eq:regul_min} is replaced with a TV prior, see \cite{Nikitin:19b}), (3) the proposed method denoted by PnP-GAN. In Rows 1--3, we demonstrate 2D slices of the 3D simulated chip to give the details of the image at different noise levels, and in the last row, we show the 3D reconstruction for the high-noise simulation. While the first row of Fig.~\ref{fig:Recons} corresponds to the well-sampled data at high-noise level, the remaining rows correspond to the under-sampled data at two different noise levels. While most of the features are recovered with well-sampled data using a sparse prior such as TV, the reconstructions are blurred. We observe that PnP-GAN not only removes the artifacts generated by ML, but also denoises images without the blurring effect. Next, we report reconstructions with under-sampled data in Fig.~\ref{fig:Recons} to highlight the effect of the proposed method, Rows 2--4. While the features are sharper at $I=[0, 968]$, the loss of quality is clear as the noise level increases at $I=[0,123]$, see Rows 2--4. Without prior knowledge, reconstructions suffer from high noise levels as confirmed by the low PSNR values in the ML reconstructions. While using TV improves the reconstruction quality compared to the ML reconstructions, the blurring effect is still visible in all reconstructions. On the other hand, PnP-GAN improves reconstruction quality with the help of iterative denoising and generates sharp images with significantly higher PSNR values. 

\begin{figure}
     \centering
     \includegraphics[width=0.95\textwidth,]{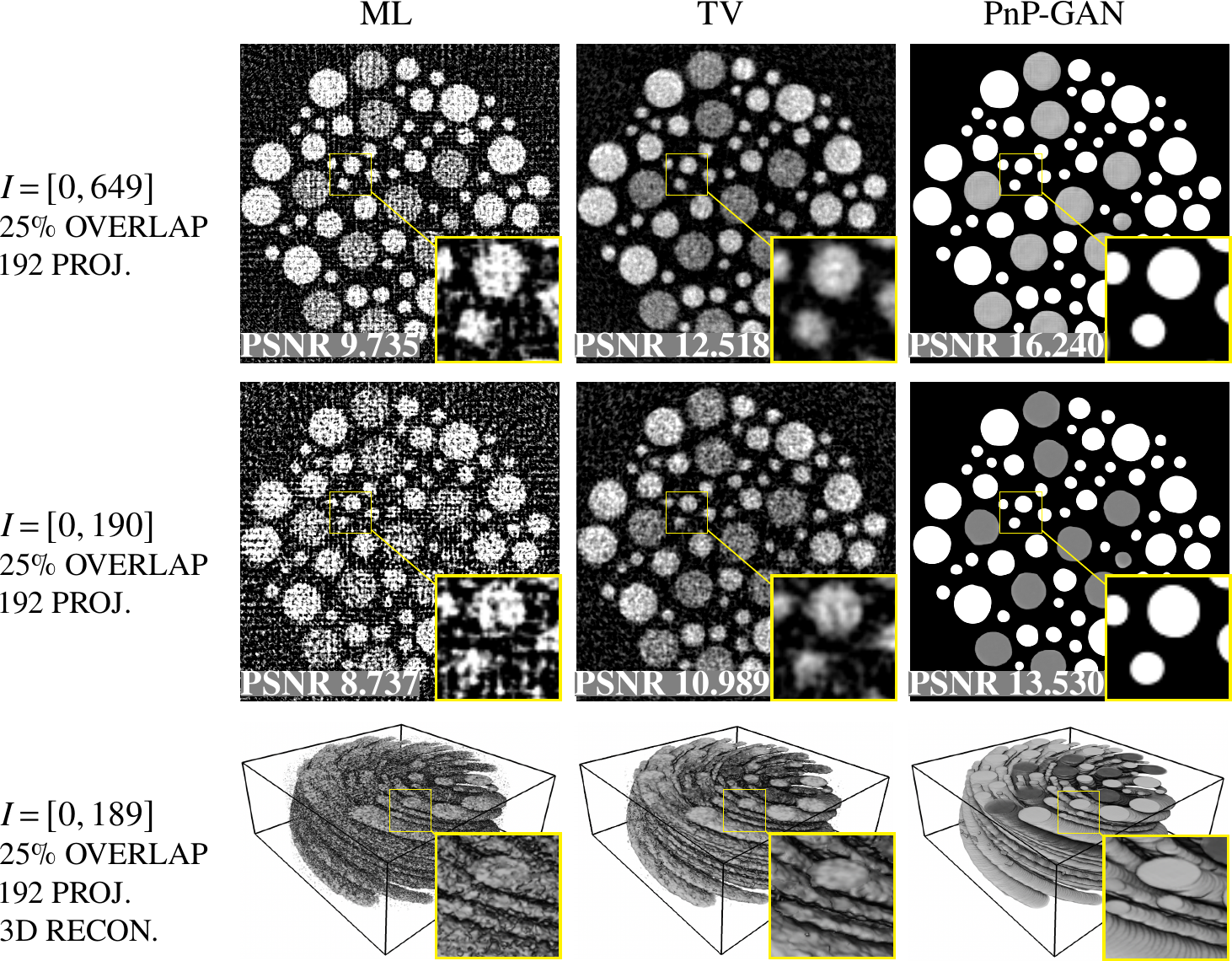}
     \caption{Reconstruction results of a slice of the 3D synthetic phantom of size $64\times 512\times 512$. We use $25\%$ probe overlap and 192 projection angles and two different noise levels. ML: the ADMM outer iteration is 250, and inner CG iterations are 4 for each ptychography and tomography subproblem. TV: the ADMM outer iteration is 250, and inner CG iterations are 4 for each ptychography and tomography subproblem followed by TV subproblem. PnP-GAN: the ADMM outer iteration is 250, and inner CG iterations are set to 4 for each ptychography and tomography subproblem followed by TomoGAN subproblem.}
     \label{fig:Recons_circles}
 \end{figure}

In this paper, our main focus is to generate good-quality reconstructions under limited and noisy measurement data. Therefore, in the next experiment, we only demonstrate reconstructions with under-sampled data where the probe overlap is $25\%$ and the object is rotated $3N/8$ times at regular intervals from $0$ to $\pi$. \selin{Reconstruction results for $\delta$ are reported in Fig.~\ref{fig:Recons_circles} using ML, MAP (with TV), and the proposed PnP-GAN methods. The first two rows of Fig.~\ref{fig:Recons_circles} demonstrate 2D slices of the 3D simulated chip to provide the details of the image at low- and high-noise levels, and in the bottom row, we show the 3D reconstruction for the high-noise simulation. The imaging artifacts in ML when there is no prior information or regularization is due to the combination of under-sampling artifacts in tomography (also known as the streaking artifacts) and measurement noise in diffracted measurements. TV regularization partly compensates for the high-frequency artifacts in images but a residual artifact pattern is still visible unless the regularization parameter is selected to be too high. PnP-GAN can successfully recover a good-quality image as demonstrated visually and through a high PSNR.}

Simulations show that the proposed method can decrease the total number of projections by 75\% based on Nyquist sampling with significantly fewer overlapped regions while generating good quality reconstructions. Although small artifacts are introduced in the reconstructions with under-sampled measurements at high noise levels, the proposed method method still gives the highest PSNR value. The reconstructions can be further improved by extending the training data or using different deep-learning-based denoisers. Our goal is not to favor a single deep-learning-based denoiser, but to introduce a generic framework that integrates such learned priors into the ADMM framework to remove the unique type of noise in ptycho-tomography problem.

\selin{To give some perspective for the computational performance of the proposed method, consider the second numerical experiment with the object size of  $180\times 512\times 512$ and detector size of $128\times 128$. We implemented the main solvers using CUDA and accelerated their computations with NVIDIA RTX 2080 GPUs. The total time for recovering the image is around 10 hours when we set the outer ADMM iterations to 250, inner CG iterations to 4 for ptychography and tomography subproblems and an additional denoiser step at each ADMM iteration.}

\begin{figure}
     \centering
     \includegraphics[width=0.4\textwidth,]{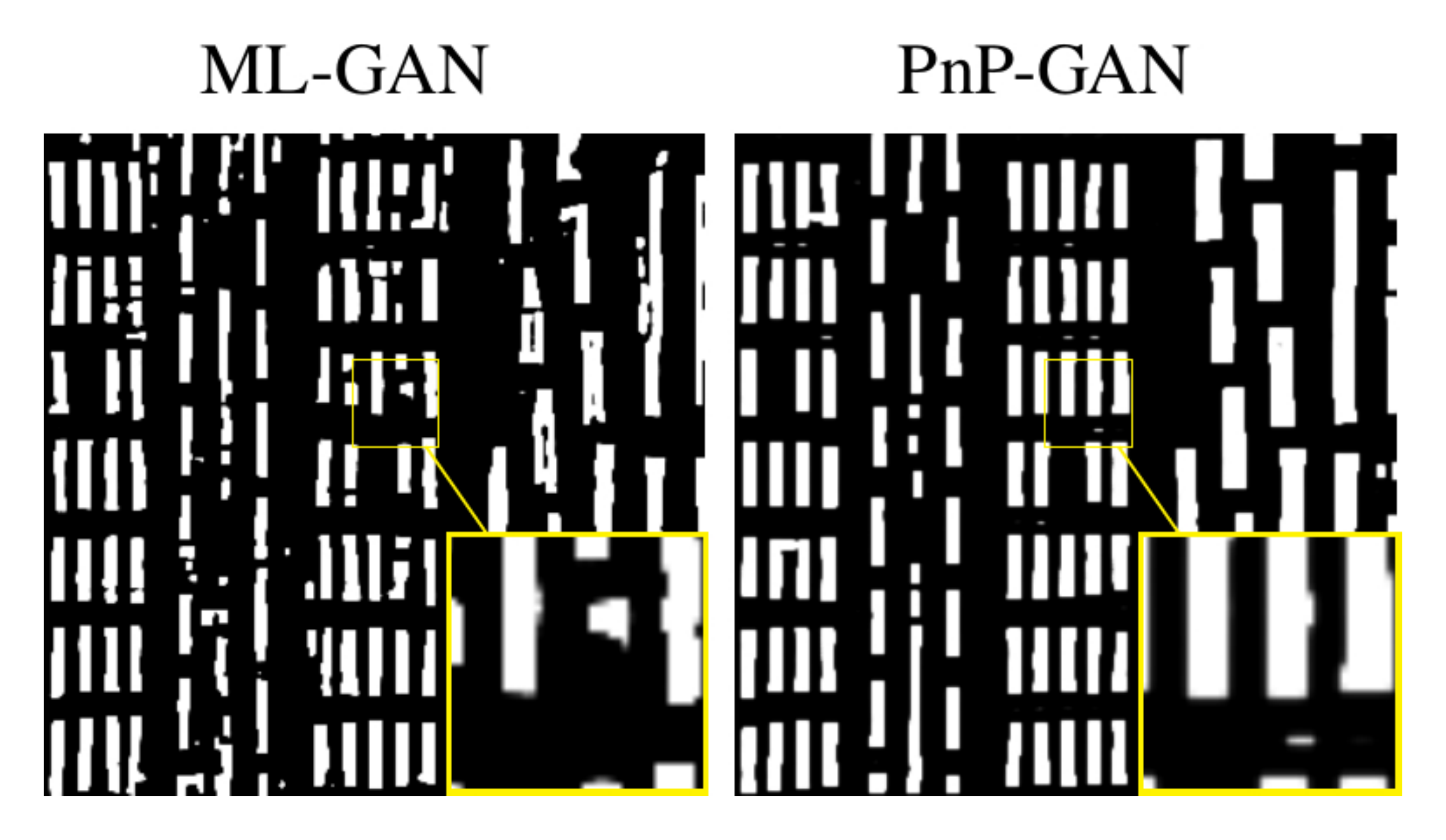}
     \caption{Reconstruction results of a slice of the 3D synthetic chip of size $64\times 512\times 512$. We use $25\%$ for $I=[0,123]$. We use $25\%$ probe overlap and 192 projection angles. ML-GAN: the ADMM outer iteration is 250, and inner CG iterations are 4 for each ptychography and tomography subproblem, and TomoGAN is applied as a postprocess. PnP-GAN: the ADMM outer iteration is 250, and inner CG iterations are set to 4 for each ptychography and tomography subproblem followed by TomoGAN subproblem.}
     \label{fig:Recons_mlgan}
 \end{figure}

In the remaining of this section, we want to show the advantage of using PnP-GAN as opposed to ML-GAN where TomoGAN is used as a postprocess denoiser. In ML-GAN, we first solve the joint ptycho-tomography problem using the ADMM method as in \cite{Aslan:19}. Then, TomoGAN is applied to the resulting reconstruction as a postprocess denoiser. To be consistent with the experiments in this paper, we also set the ADMM outer iteration to 250, and inner CG iterations to 4 for each ptychography and tomography subproblem. On the other hand, PnP-GAN splits the joint problem into three parts: ptychography, tomography and denoiser where learned priors are used for iterative denoising at each iteration of the ADMM method. \selin{The reconstruction results are shown in the Fig.~\ref{fig:Recons_mlgan}. While ML-GAN can generate decent reconstructions with well-sampled data, the reconstruction quality highly depends on the noisy input of the image. Therefore, the degradation in image quality is severe when using under-sampled and noisy measurement data. On the other hand, PnP-GAN iteratively denoises the input image, and improves reconstruction quality even at high noise levels. This is potentially due to the nature of the iterative optimization process, where earlier iterations are less noisy and a tunable GAN model is effectively enhancing the image without generating artificial artifacts. In fact, the effect of learned priors is more drastic at high noise levels because the small features in the ML estimate are hardly separable from the background.}

 \subsection{Effect of the denoiser parameter, $\alpha^k$}
In this section, we present an empirical study on the effect of the denoiser parameter, $\alpha^k$, based on reconstruction quality and residual decay using six representative schemes. The goal of this section is not to provide an optimal denoiser parameter, but to share valuable observations to decide on an effective one. 

\begin{figure}
\centering
\includegraphics[width=0.9\textwidth,]{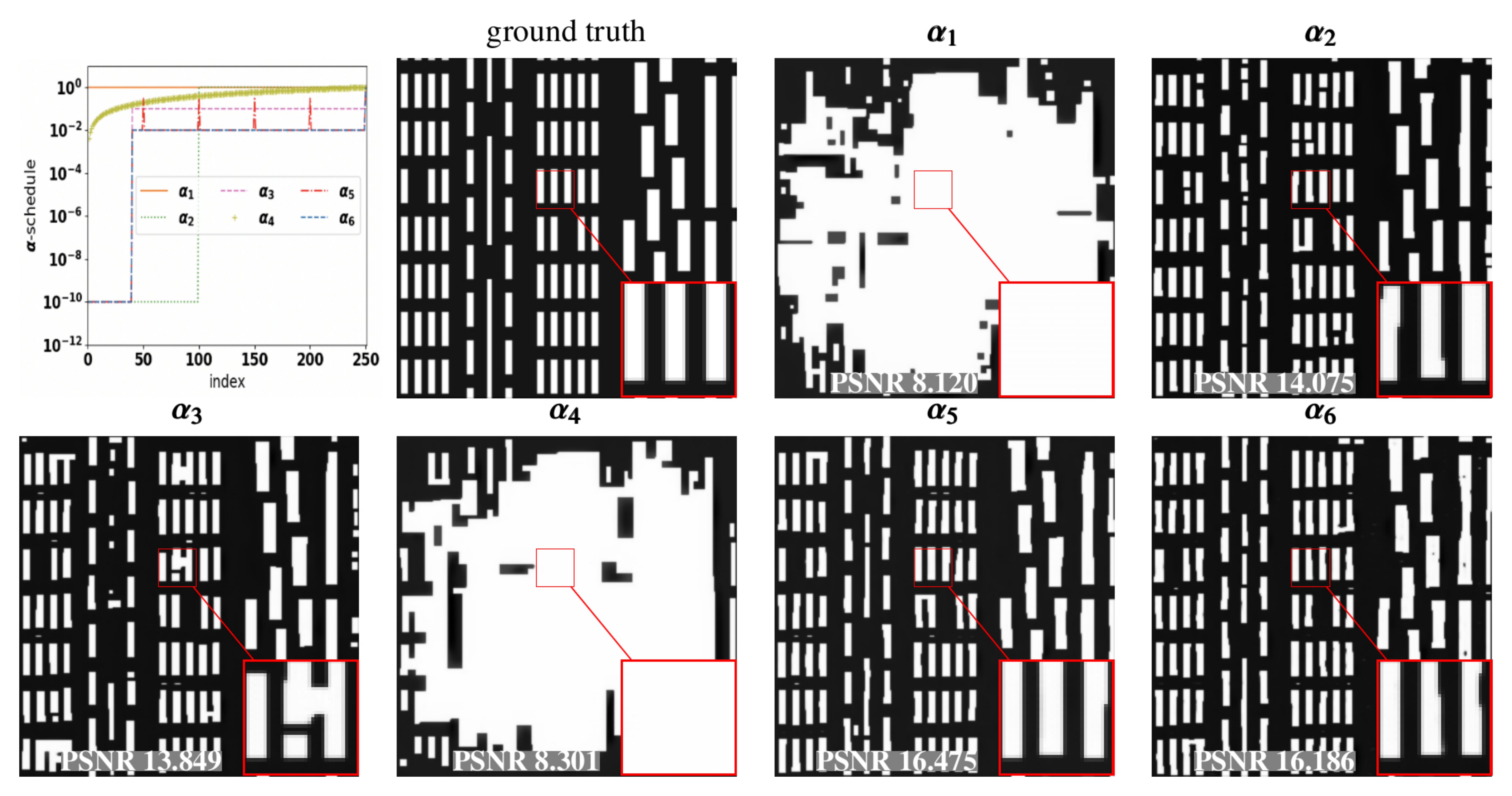}
 \caption{The reconstruction results for the corresponding $\alpha^k$ values.}
 \label{fig:TuningParamLine}
\end{figure}

In Fig.~\ref{fig:TuningParamLine}, we give the $\alpha$-schedules and reconstructions with corresponding $\alpha^k$ values. To demonstrate reconstruction quality, we provide 2D slice of the simulated chip for each denoiser parameter and report the PSNR value on each image. In some cases, we observe that MSE loss in TomoGAN causes some peak amplitude information to be lost since it tries to fit the average. However, this loss does not affect the image quality notably as it is confirmed with relatively high PSNR values. 

\begin{figure}
\centering
\includegraphics[width=0.95\textwidth,]{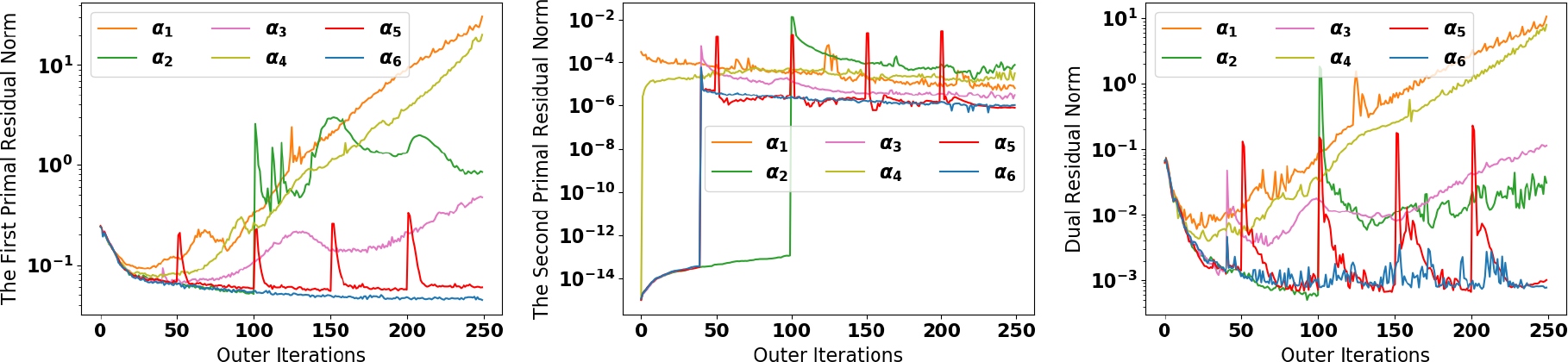}
  \caption{Residual decays for the corresponding $\alpha^k$ values.}
 \label{fig:TuningParamConv}
\end{figure}

To highlight the effect of the denoiser parameter on convergence, we also monitor the optimality conditions for the ADMM problem, which are the primal and dual feasibility. For our problem, the primal residuals for the two constraints at iteration $k+1$ are defined as follows:
\begin{align}\label{Eq:PR}
  r_1^{k+1} &= \Hop x^{k+1} -\psi^{k+1} \:\:\:\:\:\: \text{and} \:\:\:\:\:\:
  r_2^{k+1} = x^{k+1}-\eta^{k+1},
\end{align}
which we call the first and second primal residuals, respectively. In addition, we define the residual for dual feasibility at iteration $k+1$ as follows:
\begin{align}\label{Eq:DR}
  s^{k+1} &= \Hop x^{k+1} - \Hop x^{k};
\end{align}
see ~\cite[Section.~(2.3)]{Aslan:19}. In Fig.~\ref{fig:TuningParamConv}, we show the residual decays for each $\alpha_k$ values. 

To summarize, we conclude that we obtain poor reconstructions in the early ADMM iterations for the joint ptycho-tomography problem using the general PnP denoising operator in Eq.~\eqref{Eq:PnPdenoiser}, $\alpha_1$. Hence, reducing the denoiser effect is essential in the first few tens of outer ADMM iterations. Furthermore, we observe that solving $\psi$ and $x$ subproblems higher number of inner iterations does not improve the reconstruction quality in the early ADMM iterations and denoiser parameter is still needed. Next, we implement the denoising operator only incrementally and maximize the denoiser effect as a postprocessing step in the final iteration. This selection not only gives one of the highest PSNR values but also gives the fastest convergence behavior, as can be confirmed in Fig.~\ref{fig:TuningParamConv}. While $\alpha_5$ also generates good-quality reconstructions, we observe that the oscillation in $\alpha_k$ values leads to oscillation in the residual decays. Our observations show that an effective denoiser parameter satisfies the convergence criteria and produces good-quality reconstructions. We obtain reconstructions with high PSNR values in both experiments using the same denoiser parameter, $\alpha_6$. While denoiser parameter requires tuning, the observations presented in this paper are applicable to other applications as well. 

\section{Discussion and Conclusions}\label{sec:discussion}
In this paper, we derive a generic reconstruction framework for solving the joint ptycho-tomography problem with learned priors. The framework splits the joint problem into three parts: ptychography, tomography, and a learned denoiser. The PnP framework is proposed as a flexible way to add state-of-the-art priors to the ADMM. For the joint ptycho-tomography problem, however, these denoisers are not effective because of the different noise characteristics in reconstructions.  To this end, we adopted a Poisson process to accurately model our measurements, and further improve reconstruction quality with deep generative models as priors. 

A popular way to speed up the ADMM method is through early termination of the subproblems. In our previous work~\cite{Aslan:19}, we observed that by solving only a few iterations of ptychography and tomography subproblems, we obtained good-quality reconstructions. In this work, we showed that the general PnP framework leads to poor denoising visible as big white blocks in the reconstructions; see $\alpha_1$ and $\alpha_4$ in Fig.~\ref{fig:TuningParamLine}. In our simulations, we discuss the importance of the denoiser parameter and introduce an empirical way to control the denoising process. Even though an optimal selection rule is challenging because of the nonconvex nature of the problem, this empirical strategy allows to obtain good results and maintains a robust inversion.  

Another way to improve the time-to-solution performance of the ADMM method is to use high-performance many-core architectures, such as GPUs. 
Depending on the  algorithm used, the solution of each subproblems defined in our framework can require significant computational throughput~\cite{Nikitin:19b,Bicer:17}.
In our work, we implemented the main solvers using CUDA and accelerated their computations with NVIDIA RTX 2080 GPUs. Similarly, we implemented \TOMOGAN{} in TensorFlow, which can be ported to and executed on variety of GPUs, for efficient training and inference operations. \selin{While our code is not yet optimized or parallelized, the approach is scalable \cite{Nikitin:19b, Nikitin:21} and there are available software frameworks that we can translate our approach to significantly improve runtimes \cite{Yu:21}}. We plan to further improve the computational performance of our solvers and intermediate steps using the methods introduced in our previous works~\cite{hidayetouglu2019memxct, Bicer:15} and provide a comprehensive evaluation in a future work.

In this work, we focused on under-sampled and highly noisy measurements where we reduced the probe overlap to $25\%$ and projection angles to $3N/8$. Hence, we decrease the total number of projections by 75\% compared with well-sampled data. Our simulations show that we can successfully resolve features at high noise level. While a serial approach of using \TOMOGAN{} for denoising after reconstruction improves reconstruction quality at lower noise levels, the degradation in image quality is substantial at high noise. In addition, our proposed framework generates a reconstructed object with minimal loss in the quality.

\selin{We demonstrated the effectiveness of the framework using synthetic 3D images from under-sampled and noisy measurement data. It should be noted that ptycho-tomography is a relatively new 3D coherent imaging technique and instruments and thus collecting experimental data is not always available for validation studies. For example, at the Advanced Photon Source today, there is no dedicated beamline to ptycho-tomography for general users. While this situation will change soon with the upcoming upgrades of the diffraction limited storage rings at multiple light source facilities worldwide, there are other common challenges in order to work with experimental data that we have not considered in writing this manuscript. For example, because the spatial resolution is on the order of nanometers, it is challenging to precisely know the position of the measurement geometry. This is a research field by itself, and there are numerous numerical methods to estimate those positions of the data points (also known as the geometrical data alignment or probe position correction; e.g. see ~\cite{Maiden:12, Gursoy:17b} and references therein), however those are additional inverse problems and add complexity to the understanding of the learned image priors in the context of ptycho-tomography inverse problem. Naively, we think that those geometrical estimation problems can be solved by introducing an additional set of new auxiliary variables in Eq.~\ref{Eq:MAP2}. by solving those inverse problems as part of a larger joint solver, however, we leave this work as a future investigation.}

\selin{Similar to other supervised learning methods, PnP-GAN technique is only applicable when a training dataset is available. This requires either knowing of the expected structure in the images before data is acquired or collecting a representative high-quality dataset (through oversampling) for training the model. While this may not be applicable to all types of samples or specimens, there are key applications that we think will benefit from this supervised approach. For example, the blueprints of integrated circuits (i.e., the GDS file that layouts the design of the chips) could be used to train the model before the experiment, and then the model can be used as part of our framework during image reconstruction. Another potential application could be brain imaging, where the training can be performed from data collected at an electron microscope.}

\selin{Even there is training data available, it is always finite and often is not completely representative of the whole set of images that are being reconstructed. Because these types of samples have repeating components or structures, our main concern of using GANs as priors is the bias to learned patterns, because sometimes GAN can create imaginary structures (i.e. image artifacts) from arbitrary noise patterns ~\cite{2014arXiv1406.2661G}. To evaluate this effect, we performed numerical tests to evaluate this bias. We removed reasonably small wires (rectangles in Fig.~\ref{Fig:3D chip}) to break the possible learned pattern of the arrangement of wires and check if the network would add those wires back during reconstruction to preserve their arrangement. To our surprise, we haven't observed such imaginary artifacts as we may expect at high noise levels. We think this behavior could be explained by the joint (and iterative) solution strategy. Earlier iterations in tomographic reconstruction create blurry but less noisy reconstructions, therefore, the model is not affected by measurement noise as it would be affected in a single image denoising application. We think this may be one of the main advantages of our approach and we believe using GANs as part of an iterative optimization technique may be applicable to other types of imaging modalities as well. As a side note, we also observed through numerous numerical tests that scaling up or down image features doesn't affect from a potential bias to size of the structures. This is more understandable because we use data augmentation through rotation and scaling for enriching the training data. Ultimately, we conclude that this approach is potentially applicable and can provide improved results if the samples are reasonably sparse such that a well-represented training data can be generated.}

\section*{Acknowledgments}
Argonne National Laboratory’s contribution is based upon work supported by Laboratory Directed Research and Development (LDRD) funding from Argonne National Laboratory, provided by the Director, Office of Science, of the U.S. Department of Energy under Contract No. DE-AC02-06CH11357. Z. Liu was supported by the Advanced Scientific Computing Research program of U.S. Department of Energy. The data generated and/or analysed during the current study are not publicly available for legal/ethical reasons but are available from the corresponding author on reasonable request.

\bibliographystyle{abbrv}
\bibliography{refs.bib}


\end{document}